\documentclass[12pt]{article}
\usepackage{amssymb,amsmath,epsfig}

\begin{document}

\title{\bf  Effects of Electromagnetic Field on The Collapse and Expansion of Anisotropic Gravitating Source}
\author{G. Abbas \thanks{ghulamabbas@ciitsahiwal.edu.pk}
\\Department of Mathematics, COMSATS Institute\\
of Information Technology, Sahiwal-57000, Pakistan.}
\date{}
\maketitle
\begin{abstract}
This paper is devoted to study the effects of electromagnetic on the
collapse and expansion of anisotropic gravitating source. For this
purpose, we have evaluated the generating solutions of
Einstein-Maxwell field equations with spherically symmetric
anisotropic gravitating source. We found that a single function
generates the various anisotropic solutions. In this case every
generating function involves an arbitrary function of time which can
be chosen to fit several astrophysical time profiles. Two physical
phenomenon occur, one is gravitational collapse and other is the
cosmological expanding solution. In both cases electromagnetic field
effects the anisotropy of the model. For collapse the anisotropy is
increased while for expansion it deceases from maximum value to
finite positive value. In case of collapse
there exits two horizons like in case of Reissner-Nordstr$\ddot{o}$m metric.\\
\end{abstract}

\section{Introduction}

Gravitational collapse is defined as the astronomical phenomenon in
which a star contracts to a point under the effect of its own
gravity. It occurs when internal nuclear fuel of a massive star
fails to supply high pressure to balance gravity. According to
General Relativity (GR), gravitational collapse of massive objects
results to the formation of spacetime singularities in our universe
(Hawking and Ellis 1979). One of the most debatable problems in GR
is the end state of massive star, which undergoes to gravitational
collapse after exhausting its nuclear fuel. What would be the kind
of singularity (covered or naked) forming due to the gravitational
collapse? To answer this question, Penrose (1969) proposed a
hypothesis known as \textit{Cosmic Censorship Hypothesis} (CCH),
which states that the end state of gravitational collapse must be a
black Hole (BH) under some realistic conditions.

Despite of various attempts over the past four decades, this problem
remained unsolved at the foundation of BH physics. By the failure of
numerous attempts to establish the CCH, it seems natural to ask what
is really the nature of spacetime singularity? This leads to study
the dynamics of gravitational collapse in more extensive way in the
framework of GR. It is urged that the final fate of the
gravitational collapse would be BH or naked singularity (NS)
depending upon the nature of initial data of the collapse. The
existence of NS in gravitational collapse would be predicted if
there are some families of timelike geodesics which end at
singularity in the past. On the other hand, no such families of
geodesics originate from the singularity when end state of the
gravitational collapse is BH (Joshi 1993). In this case, the
spacetime singularity would be hidden by the event horizon of
gravity, while for NS there is a causal correspondence between the
region of spacetime singularity and external observers.

The gravitational collapse of dust matter was studied by Oppenheimer
and Snyder (1939) many years ago. Since dust is not a realistic
matter so more analytic analysis of collapse was made by Misner and
Sharp (1964) for perfect fluid collapse. In order to include the
effects of anisotropy in the gravitating source, Misner and Sharp
(1965) discussed the collapse of anisotropic fluid. After this there
has been a growing interest to investigate the collapse of
anisotropic stars. Many authors (Bayin 1982, Bondi 1992, Barcelo et
al. 2008, Cosenza et al. 1981 and Sharif and Abbas 2013a, 3013b,
2013c) have pointed out the applications of anisotropic solutions to
stellar collapse. An extensive study on anisotropic gravitating
source was carried out by Herrera and his collaborators (Herrera et
al. 2008a, 2008b). Several authors (Cognola et al. 2007, Gasperini
et al. 1993, Nojiri et al. 2005, Nojiri et al. 2006) have discussed
the anisotropy of dark energy in modified theories of gravity.

Also, Herrera and Santos (1997) have investigated the some
properties of anisotropic self-gravitating system and determined the
stability of the perturbed system. They proposed the possibility for
the existence of a single generated function. This method of
generating function was developed by the Glass (1981) to study the
gravitational collapse of radiating fluid. In a recent paper, the
same author (2013) has formulated a model of zero heat flux
anisotropic fluid sphere which exhibit either expansion or collapse,
depending on the choice of time profile and initial data. In a
recent paper (Abbas 2014a), we have developed the plane symmetric
model of collapse. In the present paper, we extend this work to
charged anisotropic spherical source.

The implementation of electromagnetic field in cosmological and
astrophysical processes is an attractive research area in
theoretical physics. Many investigations in this direction are
devoted to understand the interaction between electromagnetic and
gravitational fields. However, little is known about the effects of
electromagnetic field on gravitational collapse of massive objects.
Thorne (1965) studied cylindrically symmetric gravitational collapse
with magnetic field and concluded that magnetic field can prevent
the collapse of cylinder before singularity formation. Ardvan and
Partovi (1977) investigated dust solution of the field equations
with electromagnetic field and found that the electrostatic force is
balanced by gravitational force during collapse of charged dust.

Stein-Schabes (1985) investigated that charged matter collapse may
produce NS instead of BH. Germani and Tsagas (2010) discussed the
collapse of magnetized dust in Tolman-Bondi model. Recently, Herrera
and his collaborators (Herrera et al. 2011, Di Prisco et al. 2007)
have discussed the role of electromagnetic field on structure
scalars and dynamics of self-gravitating objects. Sharif and his
collaborators (Sharif and Zaeem 2012a, 2012b, Sharif and Zeeshan
2012) have extended this work for cylindrical and plane symmetries.
In recent papers, Abbas (2014b,2014c), we have studied the effects
of the charge on accretion of black hole.

This paper is devoted to study the dynamics of non-adiabatic charged
spherically symmetric gravitational collapse to see the effects of
charge on the process of collapse. The plan of the paper is the
following. In the next section, we present the charged anisotropic
source and Einstein field equations. Section \textbf{3} deals with
the generating generating solutions which represent collapse as well
as expansion. In the last section, we present the results of the
paper.

\section{Interior Matter Distribution and the Field Equations}

We take non-static spherically symmetric spacetime as an interior
metric in the co-moving coordinates in the form
\begin{eqnarray}\label{1}
ds^{2}_{-}=-X^{2}dt^{2}+Y^{2}dr^{2}+R^{2}(d{\theta}^{2}+sin^2{\theta}
d\varphi^{2}),
\end{eqnarray}
where $X$, $Y$ and $R$ are functions of $t$ and $r$. Matter under
consideration is anisotropic fluid which has zero heat flux. The
energy-momentum tensor for such a fluid is defined as
\begin{equation}\label{2}
T_{\alpha\beta}=(\mu+P_{\bot})V_{\alpha}V_{\beta}+P_{\bot}g_{\alpha\beta}
+(P_{r}-P_{\bot})\chi_{\alpha} \chi_{\beta},
\end{equation}
where $\mu,~P_{r},~P_{\bot},~V_{\alpha}$ and $\chi_{\alpha}$ are the
energy density, the radial pressure, the tangential pressure, the
four-velocity of the fluid and the unit four-vector along the radial
direction respectively. For the metric (1), $V_{\alpha}$ the
four-vector velocity and $\chi_{\alpha}$ four-vector along the
radial direction are given by
\begin{equation*}\label{3}
V^{\alpha}=X^{-1}\delta^{\alpha}_{0},\quad
\chi^{\alpha}=Y^{-1}\delta^{\alpha}_{1}
\end{equation*}
which satisfy
\begin{eqnarray*}\label{3}
V^{\alpha}V_{\alpha}=-1,\quad \chi^{\alpha}\chi_{\alpha}=1,\quad
\chi^{\alpha}V_{\alpha}=0.
\end{eqnarray*}

The expansion scalar is
\begin{equation}\label{3a}
 \Theta=\frac{1}{X}\left(\frac{2\dot{Y}}{Y}+\frac{\dot{R}}{R}\right),
\end{equation}
where dot denote differentiation with respect to $t$. We define a
dimensionless measure of anisotropy as
\begin{equation}\label{3b}
\Delta a=\frac{P_r-P_{\perp}}{P_r}.
\end{equation}

We can write the electromagnetic energy-momentum tensor in the form
\begin{equation}\label{3}
T^{(em)}_{\alpha\beta}=\frac{1}{4\pi}\left(F_{\alpha}^{\gamma}F_{\beta\gamma}
-\frac{1}{4}F^{\gamma\delta}F_{\gamma\delta}g_{\alpha\beta}\right).
\end{equation}
The Maxwell equations are given by
\begin{eqnarray}\label{4}
F_{\alpha\beta}&=&\phi_{\beta,\alpha}-\phi_{\alpha,\beta},\\\label{5}
{F^{\alpha\beta}}_{;\beta}&=&4\pi J^{\alpha},
\end{eqnarray}
where $F_{\alpha\beta}$ is the Maxwell field tensor, $\phi_{\alpha}$
is the four potential and $J_{\alpha}$ is the four current. Since
the charge is at rest with respect to the co-moving coordinate
system, thus the magnetic field is zero. Consequently, the four
potential and the four current will become
\begin{equation}\label{6}
\phi_{\alpha}=\phi{\delta^{0}_{\alpha}},\quad J^{\alpha}=\sigma
V^{\alpha},
\end{equation}
where $\phi=\phi(t,r)$ is an arbitrary function and
$\sigma=\sigma(t,r)$ is the charge density.

For the interior spacetime, using Eq.(\ref{6}), Maxwell field
equations take the following form
\begin{eqnarray}\label{8}
\phi''-\left(\frac{X'}{X}+\frac{Y'}{Y}-2\frac{R'}{R}\right){\phi'}&=&{4
\pi }{\sigma}XY^{2},
\\\label{9}
{\dot{\phi}}'-\left(\frac{\dot{X}}{X}+\frac{\dot{Y}}{Y}-2\frac{\dot{R}}{R}\right){\phi'}&=&0,
\end{eqnarray}
where prime represents the partial derivatives with respect to $r$.
Integration of Eq.(\ref{8}) implies that
\begin{equation}\label{10}
\phi'=\frac{sXY}{R^{2}},
\end{equation}
where $s\left(r\right)=4{\pi}{\int^{r}_{0}}\sigma YR^{2}dr$ is the
total charge inside the sphere and is the consequence of law of
conservation of charge, $J^\mu_{; \mu}=0$. Obviously Eq.(\ref{9}) is
identically satisfied by Eq.(\ref{10}).

The Einstein field equations,
$G_{\alpha\beta}=8\pi(T_{\alpha\beta}+T^{(em)}_{\alpha\beta})$, for
the metric (1) can be written as
\begin{eqnarray}\label{12}
8{\pi}(T_{00}+T^{(em)}_{00})&=&8{\pi}{\mu}X^{2}+\frac{s^{2}
{X^{2}}}{{R^{4}}}\nonumber\\
&=&\frac{\dot{R}}{R}\left(2\frac{\dot{Y}}{Y}+\frac{\dot{R}}{R}\right)
+\left(\frac{X}{Y}\right)^{2}\left(-2\frac{R''}{R}
+2\frac{Y'R'}{YR}-(\frac{R'}{R})^2\right),\nonumber\\\\
\label{13} 8{\pi}(T_{01}+T^{(em)}_{01})&=&0=\frac{2}{XY}
\left(\frac{\dot{R'}}{R}-\frac{\dot{Y}R'}{YR}-\frac{\dot{R}X'}{RX}\right),\\
\label{14} 8{\pi}(T_{11}+T^{(em)}_{11})&=&8{\pi}P_{r}Y^{2}
-\frac{s^{2}Y^{2}}{R^{4}}\nonumber\\
&=&-\left(\frac{Y}{X}\right)^{2}\left(2\frac{\ddot{R}}{R}+\left(\frac{\dot{R}}{R}\right)^{2}
-2\frac{\dot{X}\dot{R}}{XR}\right)+\left(\frac{R'}{R}\right)^{2}+2\frac{X'C'}{XR},\nonumber\\\label{15}
8{\pi}(T_{22}+E_{22})&=&8{\pi}P_{\bot}R^{2}
+\frac{s^{2}}{R^{2}}\nonumber\\
&=&-\left(\frac{R}{X}\right)^{2}\left(\frac{\ddot{Y}}
{Y}+\frac{\ddot{R}}{R}-\frac{\dot{X}}{X}\left(\frac{\dot{Y}}{Y}+\frac{\dot{R}}{R}\right)
+\frac{\dot{Y}\dot{R}}{YR}\right)\nonumber\\
&+&\left(\frac{R}{Y}\right)^{2}\left(\frac{X''}
{X}+\frac{R''}{R}-\frac{X'}{X}\left(\frac{Y'}{Y}-\frac{R'}{R}\right)-\frac{Y'R'}
{YR}\right).
\end{eqnarray}
The  Misner-Sharp mass function with the contribution of
electromagnetic field (Di Prisco et al. 2007) is given by
\begin{equation}\label{16}
m(t,r)=\frac{R}{2}\left(\left(\frac{\dot{R}}{X}\right)^2-\left(\frac{{R'}}{Y}\right)^2+1\right)+\frac{s^2}{2R}.
\end{equation}
The auxiliary solution of Eq.(\ref{13}) is
\begin{equation}\label{17a}
X=\frac{\dot{R}}{R^{\alpha}}\quad Y={R^{\alpha}},
\end{equation}
where $\alpha$ is arbitrary constant. In this case interior metric
(1) can be written as
\begin{equation}\label{17b}
ds^{2}_{-}=-\left(\frac{\dot{R}}{R^{\alpha}}\right)^2dt^{2}+R^{2\alpha}dr^{2}+R^{2}(d{\theta}^{2}+\sin^2{\theta}
d\varphi^{2}).
\end{equation}

Now using Eq.(\ref{17a}), we have following form of expansion scalar
\begin{equation}\label{17c}
\Theta=(2+\alpha)R^{(1-\alpha)}.
\end{equation}
For $\alpha>-2$ and $\alpha<-2$, we have expanding and collapsing
regions.

Using Eq.(\ref{12}), we have the following form of matter components

\begin{eqnarray}\label{17}
8\pi\mu +\frac{s^2{\dot{R}}^2}{R^{2\alpha+4}} &=&(1+2
\alpha)R^{2{\alpha}-2}-\frac{1}{R^{2\alpha}}\left(\frac{2R''}{R}+(1-2\alpha)\left(\frac{R'}{R}\right)^2\right)+\frac{1}{R^2},
\\\label{18}
8\pi P_{r}-\frac{s^2 {R}^{2
\alpha}}{R^{4}}&=&(1+2\alpha)R^{2{\alpha}-2}+\frac{1}{R^2}+\frac{1}{R^{2\alpha}}\left((2\alpha-1)\left(\frac{R'}{R}
\right)^2-\frac{2\dot{R}'}{\dot{R}}\frac{R'}{R}\right),
\\\nonumber
8\pi
P_{\perp}+\frac{s^2}{R^4}&=&-\alpha(1+2\alpha)R^{2{\alpha}-2}\\\nonumber
&+&\frac{1}{R^{2\alpha}}\left((1-\alpha)\frac{R''}{R}-(3\alpha-1)\frac{\dot{R}'}{\dot{R}}\left(\frac{R'}{R}
\right)+\frac{\dot{R}''}{R}+\alpha(2\alpha-1)\left(\frac{R'}{R}
\right)^2\right).\\\nonumber
\\\label{19}
\end{eqnarray}
For particular values of $R(r,t)$ and $\alpha$, we can find an
anisotropic configuration.

In this case sectional curvature mass with the contribution of
electromagnetic field given by Eq.(\ref{16}) takes the following
form
\begin{equation}\label{c1}
\frac{2m(t,r)}{R}-\frac{s^2}{R^2}-1=\left({R^{2{\alpha}}}-\frac{R'^{2}}{{R^{2\alpha}}}\right).
\end{equation}

When ${R'}=R^{2\alpha}$, there exist two trapped surfaces at $R=m
\pm \sqrt{m-s^2}$, provided $m\geq s$, thus in this case
${R'}=R^{2\alpha}$ is trapped surface condition.

Following Glass (2013) , we can find two rapping scalars which are
given by
\begin{equation}\label{c2}
\kappa_{1}=\frac{{R^{{\alpha}}}}{R}+\frac{R'}{R^{2\alpha+1}}, \quad
\kappa_{2}=\frac{{R^{{\alpha}}}}{R}-\frac{R'}{R^{2\alpha+1}}.
\end{equation}
For the trapped surface condition ${R'}=R^{2\alpha}$, we get
\begin{equation}\label{c22}
\kappa_{1}= 2R^{2\alpha-1},\quad\kappa_{2}=0.
\end{equation}
This implies that gravitational collapse leads to the formation of
two trapping surfaces at $R=m \pm \sqrt{m-s^2}$. The trapping
condition ${R'}=R^{2\alpha}$, has the integral
\begin{equation}\label{c3}
R^{(1-2\alpha)}_{trap}=(1-2\alpha)r+H(t),
\end{equation}
where $H(t)$ is arbitrary function. When the trapped condition
${R'}=R^{2\alpha}$ is applied into the Eqs.(\ref{17}), (\ref{18})
and (\ref{19}), we get
\begin{equation}\label{c4}
8\pi \mu=
R^{-2}_{trap}-\frac{s^2\dot{R^2}_{trap}}{{R^{2\alpha+4}_{trap}}},\quad
8\pi P_{r}=R^{-2}_{trap}+{s^2R^{2 \alpha-4}_{trap}},\quad8\pi
P_{\perp}=-{s^2R^{-4}_{trap}}.
\end{equation}

\section{Generating Solution}

For negative and positive values of $\alpha$, we have collapsing and
expanding solutions as follows:

\subsection{Collapse with $\alpha=-\frac{5}{2}$}
For collapse, the rate of expansion must be negative, from
Eq.(\ref{17c}), $\Theta<0$, when $\alpha<-2$, we assume that
$\alpha=-\frac{5}{2}$ and the condition ${\acute{R}}=R^{2\alpha}$,
leads to ${\acute{R}}=R^{-{5}}$, the integral of this equation is
\begin{equation}\label{c5}
R_{trap}=(6r+h_1(t))^{\frac{1}{6}},
\end{equation}
where $h_1(t)$ is arbitrary function of $t$. For
$\alpha=-\frac{5}{2}$, Eqs.(\ref{17}), (\ref{18}) and (\ref{19})
give
\begin{eqnarray}\label{c6}
8\pi\mu+s^2 {\dot{R}}^2 R
&=&-4R^{-7}-2{R^{5}}\left[\frac{2R''}{R}+3\left(\frac{R'}{R}\right)^2\right]+\frac{1}{R^2},
\\\label{c8}
8\pi
P_{r}-\frac{s^2}{R^9}&=&-4R^{-7}+\frac{1}{R^2}-2R^{5}\left[3\left(\frac{R'}{R}
\right)^2-\frac{2\dot{R}'}{\dot{R}}\frac{R'}{R}\right],
\\\nonumber
8\pi
P_{\perp}+\frac{s^2}{R^4}&=&-10R^{-7}+{R^{5}}\left[\frac{7}{2}\frac{R''}{R}+\frac{17}{2}\frac{\dot{R}'}{\dot{R}}\left(\frac{R'}{R}
\right)+\frac{\dot{R}''}{R}+15\left(\frac{R'}{R}
\right)^2\right].\\\label{c9}
\end{eqnarray}
Using Eq.(\ref{c5}) the above equations reduce to
\begin{eqnarray}\nonumber
8\pi\mu
&=&(6r+h_1)^{-1/3}-\frac{s^2}{6}{{\dot{h}}^2(6r+h_1)^{3/2}},\\\label{c10}
\\\nonumber
8\pi P_{r}&=&(6r+h_1)^{-1/3}+s^2(6r+h_1)^{-3/2},\\\label{c11}
\\\label{c12}
8\pi P_{\perp}&=&s^2(6r+h_1)^{-1/3}.
\end{eqnarray}
 The mass function
Eq.(\ref{c1}), in this case takes the following form

\begin{equation}
m_1(r,t)=\frac{R}{2}\left[\frac{s^2}{(6r+h_1(t))^{3/2}}+1+\frac{1}{{(6r+h_1(t))^{5/6}}}\left(1-\frac{1}{{(6r+h_1(t))^{10/3}}}\right)\right].
\end{equation}
The dimensionless measure of anisotropy defined by Eq.(\ref{3b}) is
\begin{equation}
\triangle
a=1+\frac{s^2(6r+h_1)^{5/6}}{(6r+h_1)^{5/6}+s^2(6r+h_1)^{-1/3}}.
\end{equation}
For constant values of $h_1$, the maximum value of $\triangle a$
occurs at the center of the sphere (at $r=0$). As $r$ increases, one
gets $\triangle a=1+\epsilon$  (where $\epsilon$ is a very small
number less than 1).
\begin{figure}
\center\epsfig{file=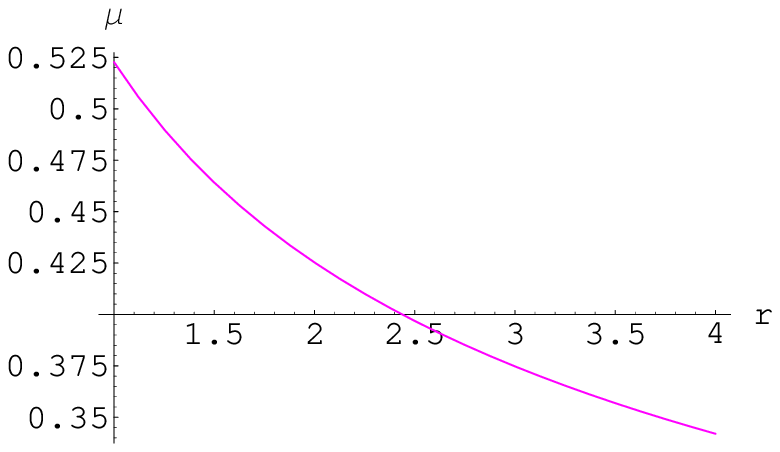, width=0.45\linewidth} \epsfig{file=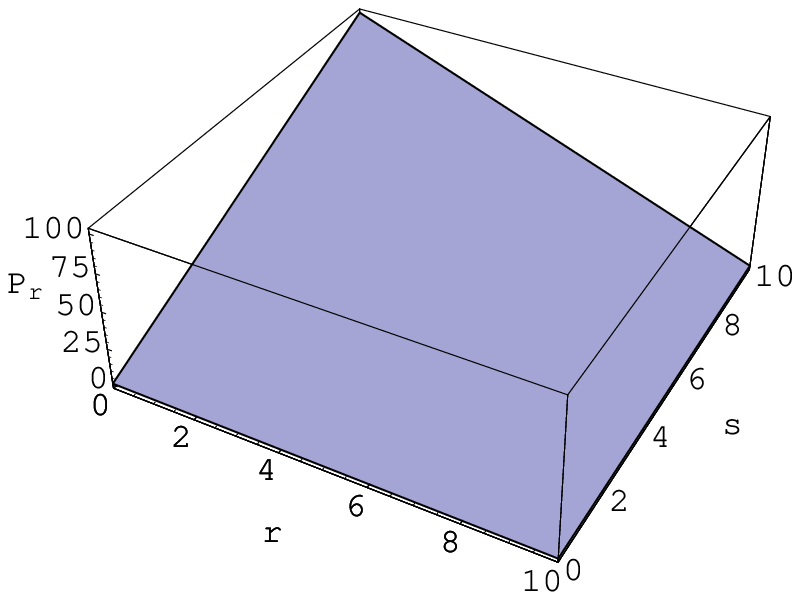,
width=0.45\linewidth}\caption{Both graphs have been plotted for
$h_1=1$.}
 \center\epsfig{file=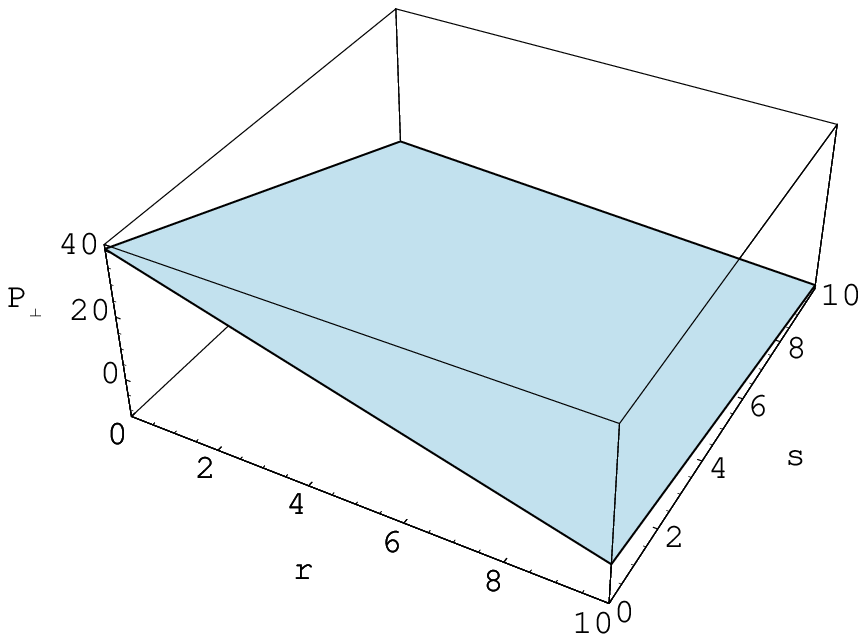,
width=0.45\linewidth} \epsfig{file=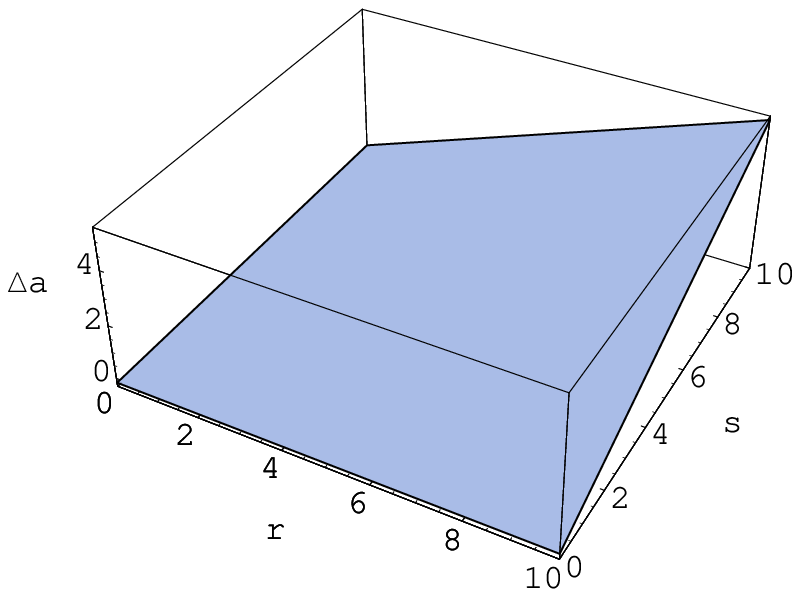,
width=0.45\linewidth}\caption{Both graphs have been plotted for
$h_1=1$.}
\end{figure}
When $\alpha $ is $\frac{-5}{2}$ then $\Theta<0$ and matter density
remains finite positive for the arbitrary choice of time profile. As
$h_1=1$ and $\dot{h_1}=0$, so the contribution of electromagnetic
field in this case vanish and density remains non-effective from
electromagnetic field. The radial and transverse pressures are
smoothly increasing and deceasing function of the electromagnetic
field, respectively. As pressure increases in radial direction and
decreases in transverse direction so anisotropy exists in the given
system. The dimensionless anisotropy is smoothly increasing function
of $s$. All these facts have been shown graphically in \textbf{1}
and \textbf{2}. The increase of anisotropy during gravitational
collapse is due to the increasing strength of electromagnetic
magnetic field. It is due to the fact that presence of
electromagnetic field in any region of spacetime causes to disturb
the generic behavior of the spacetime, so provide an external force
which enhances the anisotropy.

\subsection{Expansion with $\alpha=\frac{3}{2}$}

For expansion, the rate of expansion must be positive, from
Eq.(\ref{17c}), $\Theta>0$, when $\alpha>0$, also we assume that
\begin{equation}
R=(r^2+{r_0}^2)^{-1}+h_2(t),
\end{equation}
where $h_2(t)$ and $r_0$ are arbitrary function and constant,
respectively. For $\alpha=\frac{3}{2}$ from Eqs.(\ref{17}),
(\ref{18}) and (\ref{19}), we get

\begin{eqnarray}\label{c6}
8\pi\mu
&=&4R-2{R^{-3}}\left[\frac{R''}{R}-\left(\frac{R'}{R}\right)^2\right]+\frac{1}{R^2}-\frac{s^2{\dot{R}}^2}{R^7},
\\\label{c8}
8\pi P_{r}&=&4R+R^{-2}+2R^{-3}\left[\left(\frac{R'}{R}
\right)^2-\frac{\dot{R}'}{\dot{R}}\frac{R'}{R}\right]+\frac{s^2}{R},
\\\nonumber
8\pi
P_{\perp}&=&-6R+{R^{-3}}\left[-\frac{1}{2}\frac{R''}{R}-\frac{7}{2}\frac{\dot{R}'}{\dot{R}}\left(\frac{R'}{R}
\right)+\frac{\dot{R}''}{R}+3\left(\frac{R'}{R}
\right)^2\right]-\frac{s^2}{R^4}.\\\label{c9}
\end{eqnarray}
With $F(t,r)=1+h_2(t)(r^2+{r_0}^2)$ and $R=\frac{F}{(r^2+{r_0}^2)}$,
the density and pressures in this case are
\begin{figure}
\center\epsfig{file=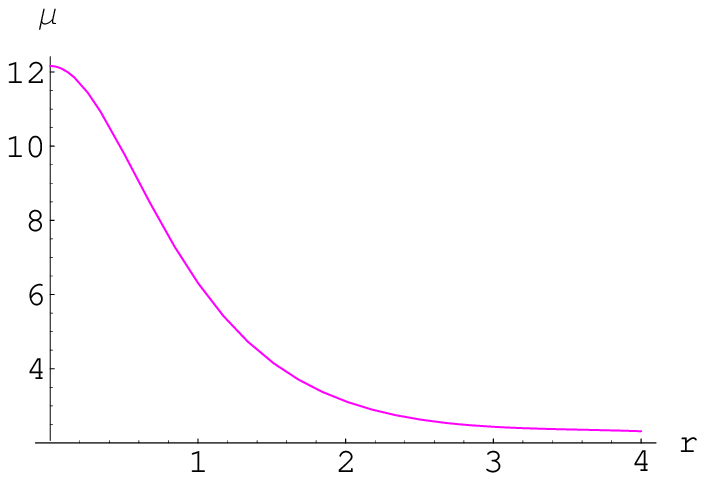, width=0.45\linewidth} \epsfig{file=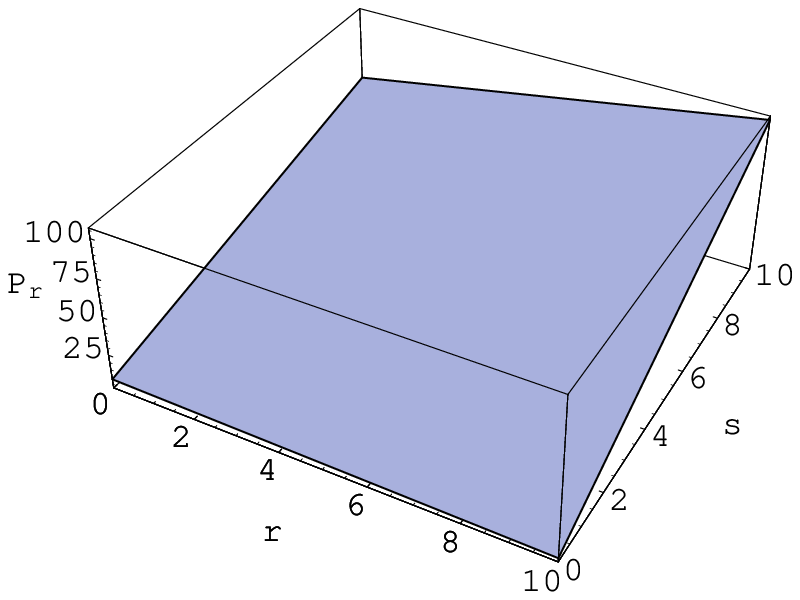,
width=0.45\linewidth}\caption{This graph is plotted for $h_2=1,
r_0=1$.}
 \center\epsfig{file=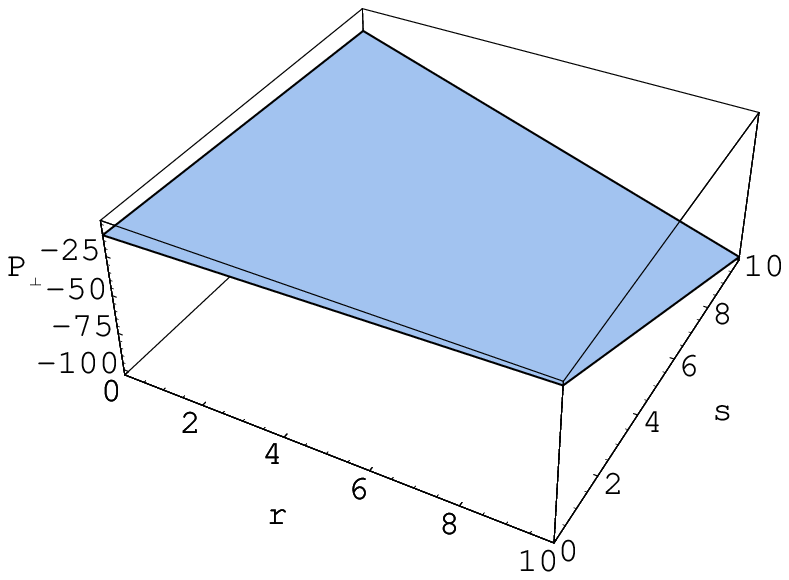,
width=0.45\linewidth} \epsfig{file=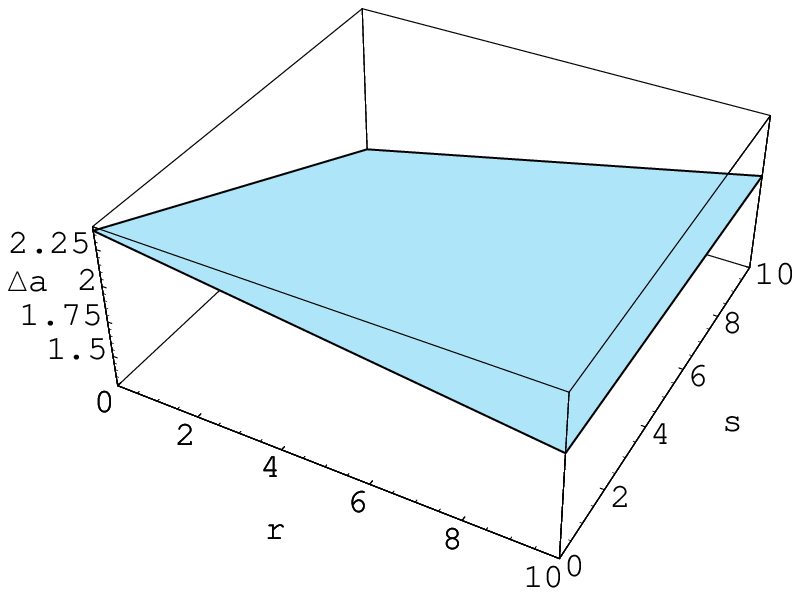,
width=0.45\linewidth}\caption{This graph is plotted for $h_2=1,
r_0=1$.} \center\epsfig{file=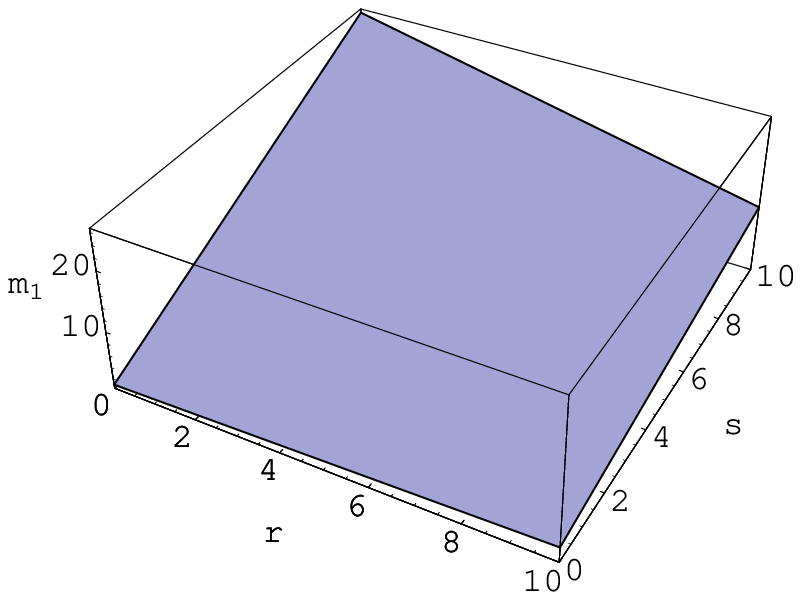, width=0.45\linewidth}
\epsfig{file=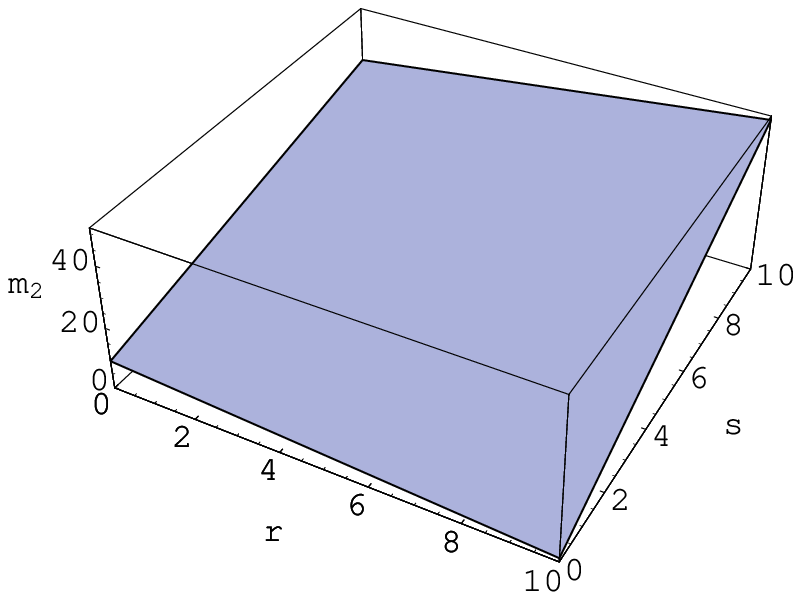, width=0.45\linewidth}\caption{This graph is
plotted for $h_2=1, r_0=1$.}
\end{figure}
\begin{eqnarray}\nonumber
8\pi\mu
&=&\frac{4F}{(r^2+{r_0}^2)}+\frac{(r^2+{r_0}^2)^2}{F^2}+\frac{4(r^2+{r_0}^2)({r_0}^2-3r^2)}{F^4}+\frac{8r^2(r^2+{r_0}^2)}{F^5}\\\label{c8}
&-&\frac{s^2{\dot{h_1}}(r^2+{r_0}^2)^7}{F^7},
\\\label{c8}
8\pi
P_{r}&=&\frac{4F}{(r^2+{r_0}^2)}+\frac{(r^2+{r_0}^2)^2}{F^2}+\frac{8r^2(r^2+{r_0}^2)}{F^5}+\frac{s^2(r^2+{r_0}^2)}{F},
\\\nonumber
8\pi
P_{\perp}&=&\frac{-6F}{(r^2+{r_0}^2)}+\frac{(r^2+{r_0}^2)({r_0}^2-3r^2)}{F^4}+\frac{12r^2(r^2+{r_0}^2)}{F^5}-\frac{s^2(r^2+{r_0}^2)^4}{F^4}.\\\label{c9}
\end{eqnarray}

The mass function Eq.(\ref{c1}), takes the following form

\begin{equation}
m_2(r,t)=\frac{F}{2(r^2+{r_0})}\left[\frac{s^2(r^2+{r_0})^2}{F^2}+1+\left(\frac{F^3}{2(r^2+{r_0})^3}-\frac{4r^2}{F^3(r^2+{r_0})}\right)\right].
\end{equation}
The dimensionless measure of anisotropy defined by Eq.(\ref{3b}) in
this case takes the following form
\begin{equation}
\triangle
a=1+\frac{\frac{6F^6}{(r^2+{r_0}^2)^2}-12r^2-F^3(r^2+{r_0}^2)+F\left[r^6s^2-{r_0}^2+3r^4s^2{r_0}^2+3r^2(1+s^2{r_0}^2)\right]}
{\left(\frac{4F^6}{(r^2+{r_0}^2)^2}+F^4s^2+8{r_0}^2+F^3(r^2+{r_0}^2)\right)}.
\end{equation}
As compared to gravitational collapse the reverse effects of
electromagnetic field occur on pressures and anisotropy during the
expansion. The expansion process causes to separate the charges
apart from each other, this leads to weak the electromagnetic field
intensity. Both the radial and transverse pressures are increasing
functions while anisotropy decreases from maximum value to finite
positive value. All these facts are shown in figures \textbf{3} and
\textbf{4}. The behavior of mass function in case of collapse and
expansion is shown in right and left graphs of figure \textbf{5},
respectively.

\section{Conclusion}

The general solution of anisotropic system has attained a
considerable interest in Einstein theory of gravity due to their
applications to stellar collapse of astrophysical objects. These
solutions helps to discuss the anisotropy of the cosmological
models. Barrow and Maartens (1998) have examined the effects of
anisotropic pressures on the late-time behavior of inhomogeneous
universe. They pointed that the decay of shear anisotropy can be
determined by measurement of anisotropic stresses. According to
Herrera and Santos (1997) in anisotropic highly dense system a phase
transition would occur during the continual collapse of the
gravitating source. They remarked that system may particularly
transit to a pion condensed state. In this case a soften equation of
state can provide a large amount of energy released during the
collapse of massive source.

This paper deals with the study of generating solution of
Einstein-Maxwell field equations with anisotropic spherically
symmetric gravitating source. Misner-Sharp mass function with the
contribution electromagnetic field as well as trapped surfaces have
been studied in detailed. The condition ${R'}=R^{2\alpha}$, implies
the existence of two horizons at $R=m \pm \sqrt{m-s^2}$, provided
$m\geq s$. This result is analogous to Reissner-Nordstrom solution
if one replaces $s=Q$. This implies that the presence of
electromagnetic field in the gravitating source is necessary and
sufficient condition for the existence of more than one horizon. In
this case $R_+$ and $R_-$ corresponds to outer and inner horizons,
respectively. The curvature singularity is hidden at the common
center of these horizons.

Under the condition ${R'}=R^{2\alpha}$, the rate of expansion scalar
becomes $\Theta=(2+\alpha)R^{(1-\alpha)}$. This implies that
$\Theta=0$, for $\alpha=-2$, $\Theta>0$, for $\alpha>-2$,
$\Theta<0$, for $\alpha<-2$, which corresponds to bouncing,
expansion and collapse, respectively. For collapse the value of
$\alpha $ is $\alpha=\frac{-5}{2}$ which implies that $\Theta<0$ and
matter density is decreasing function of $r$ for the arbitrary
choice of time profile. This becomes independent of electromagnetic
field contribution for the given time profile. The radial and
transverse pressures are increasing and deceasing function of the
electromagnetic field contribution $s$. Since pressure increases in
one direction while it increases in the other direction so the
difference of pressures is non-zero which causes the anisotropy in
the given system. The dimensionless anisotropy is increasing
function of $s$. All these facts have been shown graphically in
\textbf{1} and \textbf{2}. The increase of anisotropy in this case
is due to the increasing strength of electromagnetic magnetic field
as charged come close to each other during gravitational collapse.
It confers the fact that presence of electromagnetic field in any
region of spacetime causes to disturb the generic behavior of the
spacetime. So, increasing electromagnetic field during gravitational
collapse causes to enhance the anisotropy.

On the other hand the reverse effects of electromagnetic field occur
on pressure components and anisotropy during the expansion as
electromagnetic field is weaker in this case as compared to
collapsing case. The expansion process causes to separate the
charges apart from each other this leads to weak the electromagnetic
field. Both the radial and transverse pressures are increasing
functions while anisotropy decreases from maximum value to finite
positive value. All these facts are shown in figures \textbf{3} and
\textbf{4}. The Electromagnetic field decreases the collapse and
increases the expansion as well as bouncing by producing the
repulsive in the system.

\vspace{0.25cm}

{\bf Acknowledgment}

\vspace{0.25cm}
 We highly appreciate the fruitful comments of the anonymous referee
 for the improvements of the paper.

\end{document}